\documentclass[journal=langd5,manuscript=article]{achemso}
\setkeys{acs}{keywords = true}
\setkeys{acs}{usetitle = true}
\usepackage[version=3]{mhchem} 
\usepackage{color}
\usepackage{threeparttable}

\usepackage{lineno}
\usepackage{hyperref}
\usepackage{upgreek}

\newcommand*{\citen}[1]{%
  \begingroup
    \romannumeral-`\x 
    \setcitestyle{numbers}%
    \cite{#1}%
  \endgroup
}

\author{Jiawang Cui}
\affiliation{State Key Laboratory of Engines, Tianjin University, Tianjin, 300350, China.}
\author{Tianyou Wang}
\affiliation{State Key Laboratory of Engines, Tianjin University, Tianjin, 300350, China.}
\alsoaffiliation{National Industry-Education Platform of Energy Storage, Tianjin University, Tianjin, 300350, China.}
\author{Zhizhao Che}
\email{chezhizhao@tju.edu.cn}
\affiliation{State Key Laboratory of Engines, Tianjin University, Tianjin, 300350, China.}
\alsoaffiliation{National Industry-Education Platform of Energy Storage, Tianjin University, Tianjin, 300350, China.}
\title{Melting process of frozen sessile droplets on superhydrophobic surfaces}


\begin{document}

\begin{abstract}
Superhydrophobic surfaces can exhibit icephobicity in many ways due to their large contact angles and small rolling angles. The melting process of frozen droplets on superhydrophobic surfaces is still unclear, hindering the understanding of surface icephobicity. In this experimental study of the melting process of frozen sessile droplets on superhydrophobic surfaces, we find two types of melting morphologies with opposite vortex directions on a single-scale nano-structured (SN) superhydrophobic substrate and a hierarchical-scale micro-nano-structured (HMN) superhydrophobic substrate. Melting pattern visualizations and flow field measurements showwed Marangoni convection and natural convection occuring in the melting sessile droplets. For the HMN superhydrophobic substrate, the internal flow was found to be dominated by Marangoni convection due to the temperature gradient along the surface of the droplet. For the SN superhydrophobic substrate, Marangoni convection was inhibited by the superhydrophobic particles at the surface of the droplet, which were shed from the fragile superhydrophobic substrate during the freezing--melting process, as confirmed by surface characterizations of the substrate and flow measurements of a water pool. These results will help researchers better understand the melting process of frozen droplets and in designing novel icephobic surfaces for numerous applications.
\end{abstract}

\section{Introduction}\label{sec:1}
Droplet freezing and melting are common phase change phenomena that occur in cold conditions and play an important role in many applications, such as aerospace \cite{yamazaki21}, transportation \cite{luo18}, electricity, and energy \cite{wei19}. Studying droplet freezing and melting is important for understanding the heat transfer, fluid flow, and phase changes that occur during such processes; it is also useful for developing effective strategies for controlling the droplet freezing and melting processes and mitigating the damage caused by such processes in numerous applications.

Studies have found that superhydrophobic surfaces can exhibit different degrees of anti-icing/deicing performance in cold conditions due to their large contact angle and small rolling angle. In terms of anti-icing, condensed droplets on superhydrophobic surfaces can easily slide off the surface by gravity or bound off the surface by merging without human intervention \cite{boreyko09, feng12-frAQB, liu21DropletJump, wen14, wen17}. Chu et al. \cite{chu17} conducted an ice growth experiment of condensed droplets on three kinds of surfaces: hydrophilic surface, hydrophobic surface, and superhydrophobic surface, and found that the superhydrophobic characteristics greatly reduced droplet adhesion, thus only a small amount of ice cap could be formed at low temperatures. When droplets are attached to the surfaces, it is ideal to get enough time to remove them before freezing. Alizadeh et al. \cite{alizadeh12} found that superhydrophobic surfaces generally have longer supercooling time than hydrophilic and hydrophobic surfaces, indicating that ice nucleation was more difficult to occur on superhydrophobic surfaces at low temperatures. Their comparison in nucleation rate also showed that as the contact angle increases, the ice nuclei need to overcome a higher energy barrier to achieve spontaneous ice growth. Although Jung et al. \cite{jung11} and Nosonovsky et al. \cite{nosonovsky12} found that superhydrophobic surfaces cannot always prolong the supercooling time due to the convex surface with micro/nano-scale which may reduce the nucleating energy barrier, superhydrophobic surfaces with reasonable material selection and structural design still show their important role in anti-icing. Once ice nuclei are formed, droplet icing will be imperative. Cabello et al. \cite{montes21} and Shi et al. \cite{shi22} compared the time required for droplets to completely freeze under different contact angle conditions. Their results showed that the contact angle indirectly affects the heat transfer intensity between the droplet and the substrate, resulting in a longer time required to complete the freeze as the contact angle increases.

Delaying the freezing process does not mean permanently preventing droplets from icing. Frost can still propagate on the surface and cause droplets to freeze due to edge defects \cite{zhang23}. Superhydrophobic surfaces can not only play an anti-icing role before droplets completely freeze but also be conducive to removing the ice cap easily \cite{stamatopoulos17}. Wang et al. \cite{wang13} studied the adhesion strength of frozen droplets on different structural surfaces with the mechanical stripping method. Their results of the tensile strength of ice adhesion showed that droplets were easier to fall off with the increase of contact angle, but there were also cases where the superhydrophobic surface adhered more firmly than the hydrophobic surface. Therefore, whether the hydrophobic and superhydrophobic structures are beneficial to reducing the adhesion strength of the ice cap is still controversial \cite{bharathidasan14, chen12, davis14}. For droplets melting on surfaces, superhydrophobicity can play a vital role in the deicing process. Murphy et al. \cite{murphy17} experimentally studied a series of self-propelled behaviors of droplets during defrosting. Due to the deformation of droplets, the released surface energy could promote droplets with a large surface-to-volume ratio to roll or jump easily, which enabled the droplets to automatically detach from the surface. Under external intervention, such as changing the substrate inclination angle \cite{boreyko13, chen21}, purging with the wind \cite{zhu17}, or using structurally designed substrates \cite{huang20, liu21}, this effect can be enhanced. The superhydrophobic surface greatly promotes the movement of melting droplets and even makes them fall off before completely melting, which effectively maintains the substrate clean.

Previous studies have shown the feasibility of superhydrophobic surfaces in improving anti-icing and deicing from different perspectives. However, less attention has been focused on the melting process of frozen droplets under heating conditions and the effect of superhydrophobic surfaces on the melting process. In this study, experiments are carried out to explore the melting process of frozen sessile droplets on superhydrophobic surfaces. The melting morphology and melting time of frozen on two types of superhydrophobic surfaces with different micro/nano-structures are compared, namely single-scale nano-structured superhydrophobic substrate and hierarchical-scale micro-nano-structured superhydrophobic substrate. We find that during the melting process, the melted fluid is not stationary but exhibits strong flow vortices. The mechanisms for different melting processes on the two types of superhydrophobic surfaces are investigated by optical measurement of the internal flow fields during the melting process, theoretical comparison of the forces for Marangoni convection and natural convection, substrate surface characterization before the freezing and after the melting, and velocity field measurements for surface flow inhibition by superhydrophobic particles.

\section{Material and methods}\label{sec:2}
\subsection{Experimental setup and procedure}\label{sec:2.1}
\begin{figure}
  \centering
  \includegraphics[width=0.8\columnwidth]{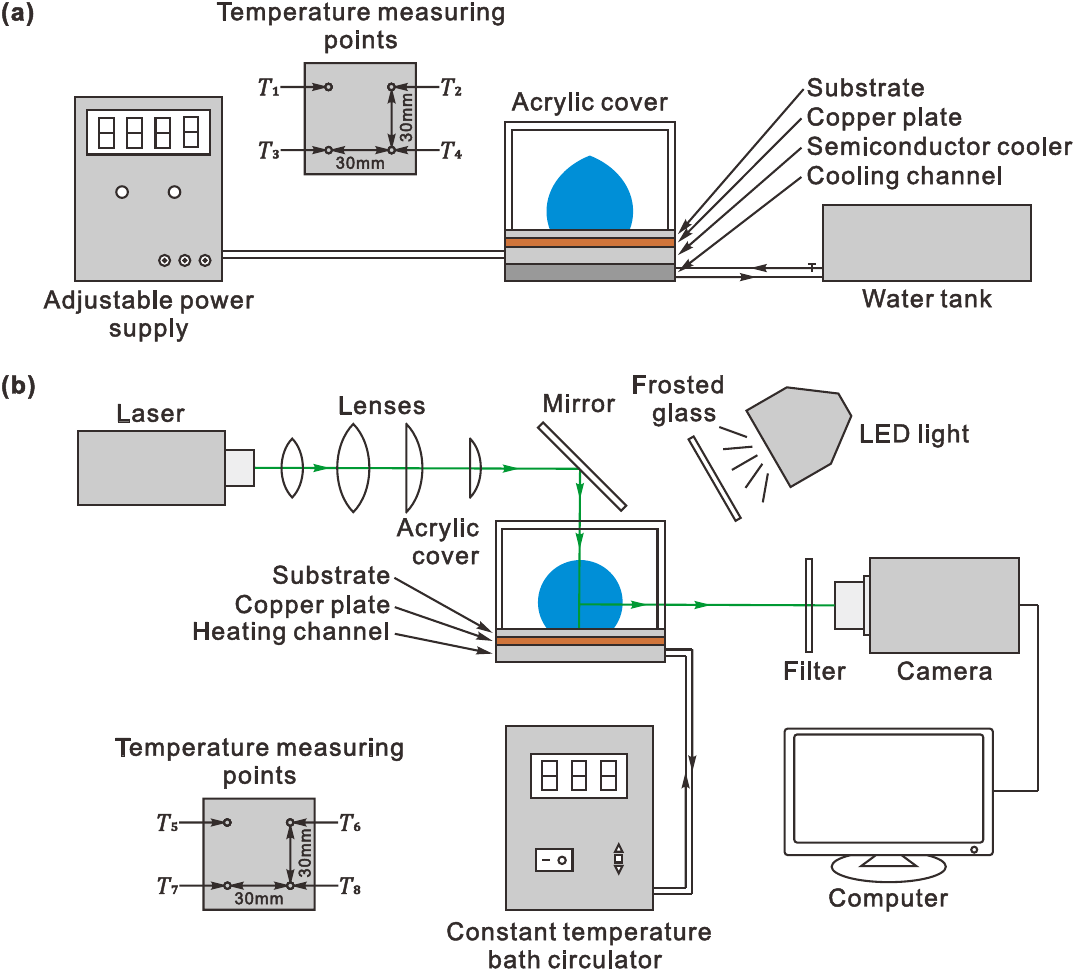}
  \caption{Schematic diagram of the experimental setup: (a) freezing part; (b) melting part.}\label{fig:01}
\end{figure}

The experimental setup consists of a freezing part and a melting part, as shown in Figures \ref{fig:01}(a) and \ref{fig:01}(b), respectively. The freezing part was used to freeze droplets on superhydrophobic substrates, and the melting part was to control and observe the melting process of frozen droplets. In the freezing part, the substrate temperature was controlled by a semiconductor refrigeration system, in which, the voltage of the DC power supply and the water flow rate in the cooling channel were adjusted according to the experimental requirement. In the melting part, the melting process was achieved by heating the substrate, and the temperature was controlled by adjusting the power of electric heating of the oil bath system. To increase the temperature uniformity of superhydrophobic substrates, a copper plate ($50 \times 50 \times 3$ mm$^3$) was set on the semiconductor cooler in the freezing part and also on the heating channel in the melting part. Four K-type thermocouples were embedded on the side of each copper plate with a spacing of 30 mm to monitor the heating temperature change. Thermal grease was added at the contact gap to improve heat conduction efficiency. During the experiment, the laboratory temperature and humidity were maintained at 23--25 $^\circ$C and 40--50\%, and an acrylic cover was used to reduce the influence of environmental fluctuations and to maintain a stable environment for the freezing and melting processes.

In the experiment, the superhydrophobic substrate was first set on the copper plate of the freezing part. The temperature of the plate was then reduced to -20 $^\circ$C by the refrigeration system. Then a droplet was deposited on the superhydrophobic substrate by releasing from a needle 8 mm above the surface. The droplet size was varied by using different needles and the droplet volume was obtained by processing the side-view images of the liquid droplet. For the melting study, the frozen droplet together with the superhydrophobic substrate was transferred to the heating part on the heating copper plate. The temperature of the heating cooper plate was increased to a prescribed heating temperature before setting the superhydrophobic substrate with frozen droplets. Because the silicon wafer is very thin (0.7 mm) and has a high thermal conductivity (148 W/(m$\cdot$K)), the superhydrophobic surface could quickly reach the heating temperature of the copper plate and induce the melting of the droplet.

The droplet morphology during the ice melting process was recorded by a CMOS camera (FLIR BFS-U3-17S7M-C) with a macro lens (TOKINA AT-X PRO 100mm), and an LED light source was used for illumination. To study the flow during the melting process, the Particle Image Velocimetry (PIV) method was used to capture the convective flow inside droplets. In flow velocity measurements, fluorescent particles (polystyrene, Thermo Fisher R0300, average diameter 3 $\upmu$m) were uniformly dispersed in droplets at a concentration of about $7.6 \times 10^3$/$\upmu$l. The effect of particles on the melting process was tested experimentally by comparing the melting morphology and the melting time, and the results showed negligible influence on the melting process. A continuum laser (CNI MGL-F-532nm, wavelength 532 nm) with a series of lenses was used to generate a sheet of light to excite fluorescent particles only at the middle cross-section of the droplet. Because of the lower energy of the laser power reaching the droplet and the transparency of the droplet to the laser wavelength, the heating effect of the laser sheet is expected to be minimal. This could also be confirmed by the unbiased shape of the unmelted ice cap of the droplet even though the laser was set on one side of the droplet. With a long-pass filter (Newport HPD550) to remove the scattering light of the laser, the fluorescent signal of the particles was recorded on the high-resolution camera. Then, the velocity field inside the droplet was obtained from the particle images (1600 $\times$ 1100 pixels) by cross-correlation algorithm using the PIV algorithm (PIV-lab). The interrogation areas of Pass 1--4 were 64 $\times$ 64, 32 $\times$ 32, 16 $\times$ 16, and 8 $\times$ 8 pixels respectively, and overlapped 50\%.

\subsection{Superhydrophobic surfaces}\label{sec:2.2}
Superhydrophobic surfaces were prepared by constructing micro/nano-scale structures on the substrates. More than 10 kinds of superhydrophobic surfaces were used to compare and analyze the melting characteristics of sessile droplets. The experimental results showed that the melting process on these superhydrophobic surfaces can be classified into two categories. Therefore, for conciseness, we selected two representative and comparable superhydrophobic substrates to describe the melting phenomenon. The melting processes on other superhydrophobic surfaces are provided in the Supplementary Material.

\begin{figure}
  \centering
  \includegraphics[width=0.7\columnwidth]{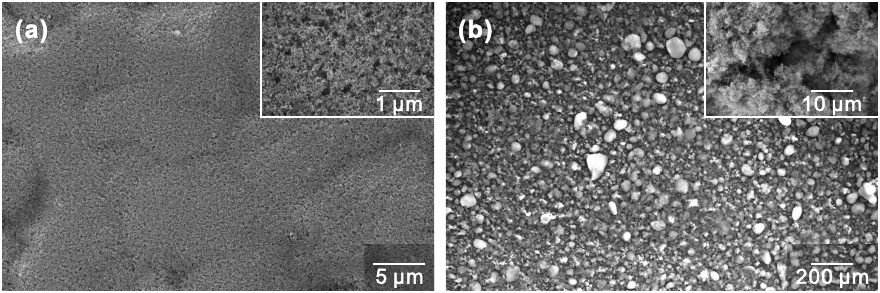}
  \caption{SEM images of the two superhydrophobic substrates: (a) single-scale nano-structured (SN) superhydrophobic substrate; (b) hierarchical-scale micro-nano-structured (HMN) superhydrophobic substrate.}\label{fig:02}
\end{figure}

Two superhydrophobic surfaces with similar chemical compositions were selected to explore the ice melting process. Silicon wafer with a size of $40\times40\times0.7$ mm$^3$ was selected as the bare substrate due to its good flatness and high thermal conductivity. One superhydrophobic substrate was to evenly spray a mixed solution (composed of 85\% isopropanol, 12\% liquefied petroleum gas, and 3\% hydrophobic silica powder) on the silicon wafer. After drying, silica particles were stacked on the substrate surface. By encapsulating air between the droplet and the substrate, the surface is superhydrophobic with a static contact angle of 151$^\circ$. The other superhydrophobic substrate was prepared by evenly coating PDMS on the silicon wafer, and then superhydrophobic silica powder was distributed on the substrate surface and fixed by PDMS evenly through vibration. After heating and solidification, the surface had a static contact angle of 153$^\circ$. Since the chemical compositions of these two substrates are similar, they can be distinguished by their micro/nano-scale structures, as shown in the SEM images in Figure \ref{fig:02}. The first type was named single-scale nano-structured (SN) superhydrophobic substrate, and the second type was named hierarchical-scale micro-nano-structured (HMN) superhydrophobic substrate.

\section{Results and discussion}\label{sec:3}
\subsection{Melting process}\label{sec:3.1}
\begin{figure}
  \centering
  \includegraphics[width=0.9\columnwidth]{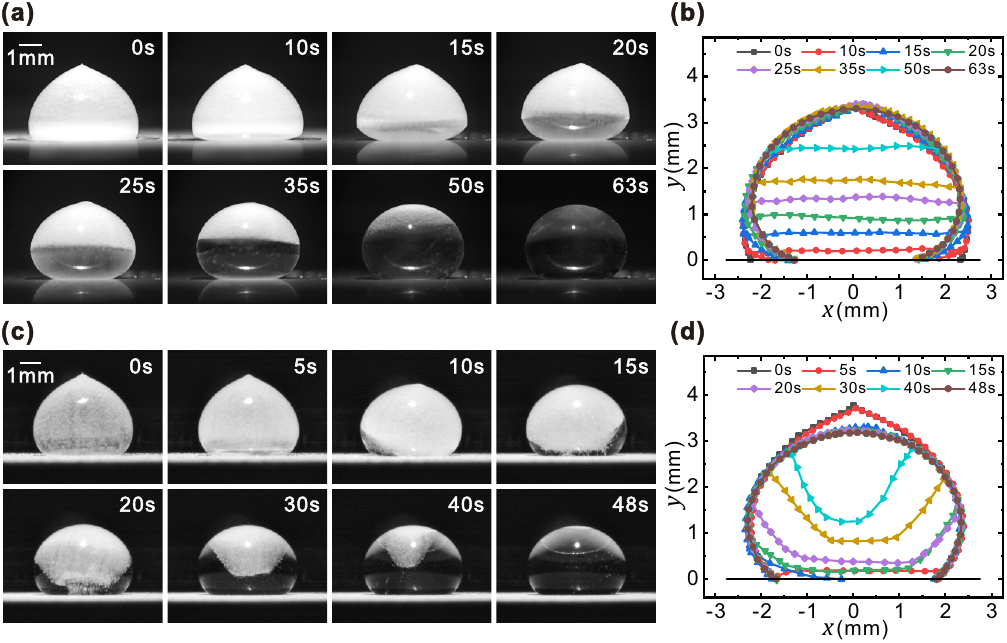}
  \caption{Morphologies of droplets and their internal ice caps during the melting processes: (a,c) melting morphologies on the SN (a) and HMN (c) superhydrophobic substrates; (b,d) contour lines on the SN (b) and HMN (d) superhydrophobic substrates. The heating temperature (i.e., the temperature of the heating copper plate in Figure 1(b)) is 30 $^\circ$C, and the droplet volume is 37.1 $\upmu$l. Video clips for the melting process are available as Supplementary Material.}\label{fig:03}
\end{figure}

The melting processes of frozen droplets on the SN and HMN superhydrophobic substrates are shown in Figure \ref{fig:03}. Before the melting stage on the SN superhydrophobic substrate, the contact angle between the frozen droplet and the superhydrophobic substrate (117$^\circ$ in Figure \ref{fig:03}(a)) is much smaller than the equilibrium contact angle of the water droplet on the substrate, which is 151$^\circ$. This phenomenon is due to the low temperature of the substrate when the droplet is placed on the SN superhydrophobic substrate: crystal nuclei have been formed during the spreading stage of the droplet and caused the bottom of the spreading droplet to freeze rapidly, thereby inhibiting the droplet retraction. Therefore, during the initial melting stage, the bottom contact region of the melting droplet contracts significantly and returns to the superhydrophobic state with a contact angle of 162$^\circ$, which is also different from the state before freezing with a contact angle of 151$^\circ$. It is obvious that the freezing and melting cycle improves the contact state of droplets, and may be useful to promote the self-separation of droplets from the substrate. A reasonable explanation for this change in the contact angle can be analyzed from two aspects. During the spreading process, the kinetic energy of the falling droplet is converted into surface energy for temporary storage, but the rapid freezing of the bottom inhibits the droplet contract. When the frozen droplet melts, the stored surface energy is released and converted into kinetic energy, which is used for the recovery of the droplet morphology. This phenomenon is helpful to gradually detach the part of the droplet pinned in the substrate micro/nano-scale structures \cite{tang2023new1, lathia2023new2}. In addition, the gas dissolved in the droplet may form bubbles and be bound in the ice cap during the freezing process \cite{li22, Chu2019new3}. According to the study of Wang et al. \cite{wang22}, these bubbles could continuously impact the substrate surface, thus providing the source of airbags to promote the detachment of the bottom of the droplet during melting. Therefore, the whole process restores the superhydrophobicity of the substrate. In contrast, on the HMN superhydrophobic substrate, there are no obvious changes in the contact line and contact angle, which means that the icing process starts after the shape of the droplet becomes stable on the superhydrophobic substrate. Even so, the contact angle during the melting process is always smaller than the static contact angle, which indicates that there should be a degradation of the surface superhydrophobicity due to the condensation \cite{feng12}. In general, the morphology difference between the SN and HMN superhydrophobic substrates indicates that superhydrophobic surfaces with similar chemical composition and contact angle may have great differences in nucleation under the same low-temperature environment, resulting in different freezing and melting characteristics.

The frozen droplets gradually melt upward from the contact region with the two superhydrophobic substrates, and the unmelted part always floats on the upper part of the droplet \cite{chu2019new4}. However, the droplet morphology during the melting process still shows great differences. On the SN superhydrophobic substrate (as shown in Figure \ref{fig:03}(a)), the unmelted ice cap has a gradually melting tip bulge due to the very slow heat exchange with the environment and its bottom has an approximately horizontal edge. When the heating temperature is higher, the bottom of the unmelted ice cap becomes concave, which means that the middle part melts faster. During the whole melting process on the SN superhydrophobic substrate, the position of the unmelted ice cap is relatively stable. As for the melting process on the HMN superhydrophobic substrate shown in Figure \ref{fig:03}(c), the top of the unmelted ice cap is smooth after a relatively short time. The disappearance of the tip bulge is because the frozen droplet preferentially melts on both sides and the bottom of the unmelted ice cap gradually becomes convex. This shape can lead to the instability of the ice cap center, which finally results in the rotation of the tip of the ice cap into the interior of the droplet. During the whole melting process on the HMN superhydrophobic substrate, the unmelted ice cap has a large shaking or rotation, especially in the early melting stage.

\begin{figure}
  \centering
  \includegraphics[width=0.5\columnwidth]{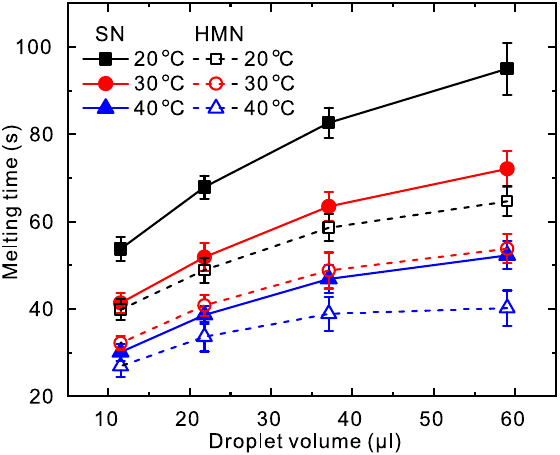}
  \caption{Melting time of frozen droplets on the SN and HMN superhydrophobic substrates under different heating temperatures and droplet volumes.}\label{fig:04}
\end{figure}

The melting time of frozen droplets on the SN and HMN superhydrophobic substrates is shown in Figure \ref{fig:04}. Time statistics start from the moment when the frozen droplets begin to melt ($t = 0$), which can be observed near the three-phase contact line. With the decrease of droplet volume and the increase of heating temperature, the ice melting time decreases on both superhydrophobic substrates, but the ice melting time on the HMN superhydrophobic substrate is always less than that on the SN superhydrophobic substrate under the same conditions. For the case of the droplet with 37.1 $\upmu$l at 30 $^\circ$C, the average melting time on the SN superhydrophobic substrate is $63.5 \pm 3.4$ s, which is in sharp contrast with the average melting time of $48.8 \pm 4.1$ s on the HMN superhydrophobic substrate. As the volume of the droplet increases, the gravitational force does alter the morphology of the frozen droplet. However, the experimental result shows that the morphology does not affect the melting mode of frozen droplets, including the melting morphology and melting flow pattern. In general, under heating conditions, the HMN superhydrophobic substrate can make the ice melt faster and greatly reduce energy consumption, which exhibits better icephobicity than the SN superhydrophobic substrate. This difference can be attributed to the melting flow inside the droplet, which can enhance heat transfer. The details of the melting flow will be discussed in the next section.

\subsection{Internal flow during the melting process}\label{sec:3.2}
Besides the melting morphology and melting time of frozen droplets, we also observed that during the melting process, the melted fluid is not stationary but exhibits strong flow vortices. The internal flow patterns are significantly different for the two superhydrophobic substrates, which may be the reason for the differences in the melting morphology and melting time. Hence, the flow inside the droplet during the melting process was measured and compared between the two superhydrophobic substrates.

\begin{figure}
  \centering
  \includegraphics{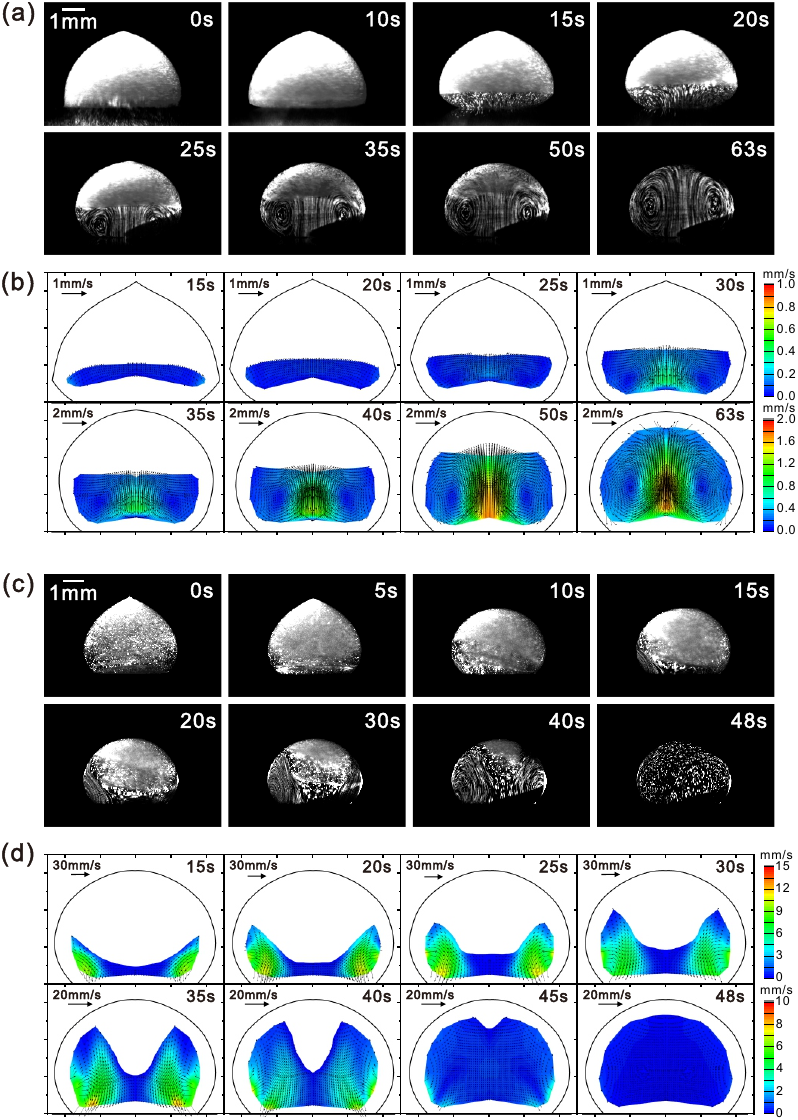}
  \caption{Internal flow of melting droplets: (a, b) raw images and velocity field of convection flow inside melting droplets on the SN superhydrophobic substrate; (c, d) raw images and velocity field of convection flow inside melting droplets on the HMN superhydrophobic substrate. The heating temperature is 30 $^\circ$C, and the droplet volume is 37.1 $\upmu$l. The velocity on the left half of the droplets in the images was mirrored to the right half because the fluorescent particles on the right half were not totally illuminated in the PIV measurement.}\label{fig:05}
\end{figure}

During the melting process, strong vortices are formed within the droplet, as shown in Figures \ref{fig:05}(a) and \ref{fig:05}(c). To obtain the velocity field quantitatively, PIV analysis was performed on the particle image sequence according to the cross-correlation method. It should be noted that the images of the particles are distorted due to the refraction of light at the gas-liquid interface of the droplet. Hence, the flow field obtained directly from the cross-correlation algorithm is incorrect. According to the geometric relation of light transmission in the PIV measurement, a mapping function between the points on the image and the droplet can be derived. The velocity is the derivative of the position over time, so the corrected velocity vectors can also be obtained by a direct mapping of the velocity vectors obtained based on the images of the droplet. The calibration process only requires optical geometric conditions, and more details can be found in Ref.~\citen{kang04}. The corrected positions and velocity vectors are shown in Figures \ref{fig:05}(b) and \ref{fig:05}(d). The velocity fields show that the internal flow patterns on the SN and HMN superhydrophobic substrates present two symmetrical vortices on both sides. As the melting proceeds, the size of the vortices increases. For the internal flow on the SN superhydrophobic substrate, the flow direction is clockwise on the right and the velocity is fast in the middle and slow on the periphery. In contrast, for the HMN superhydrophobic substrate, the internal flow shows an opposite trend, i.e., the flow direction on the right is counterclockwise and the velocity is the smallest in the middle and the largest on the periphery. With the increase in the flow intensity, the heat transfer can be enhanced, thus promoting the melting in the middle on the SN superhydrophobic substrate whereas on the periphery for the HMN superhydrophobic substrate. The characteristics of the velocity field are consistent with the change of the droplet morphologies during the melting process. This result suggests that it is the difference in melting flow that leads to the difference in melting morphology.

\begin{figure}
  \centering
  \includegraphics[width=0.8\columnwidth]{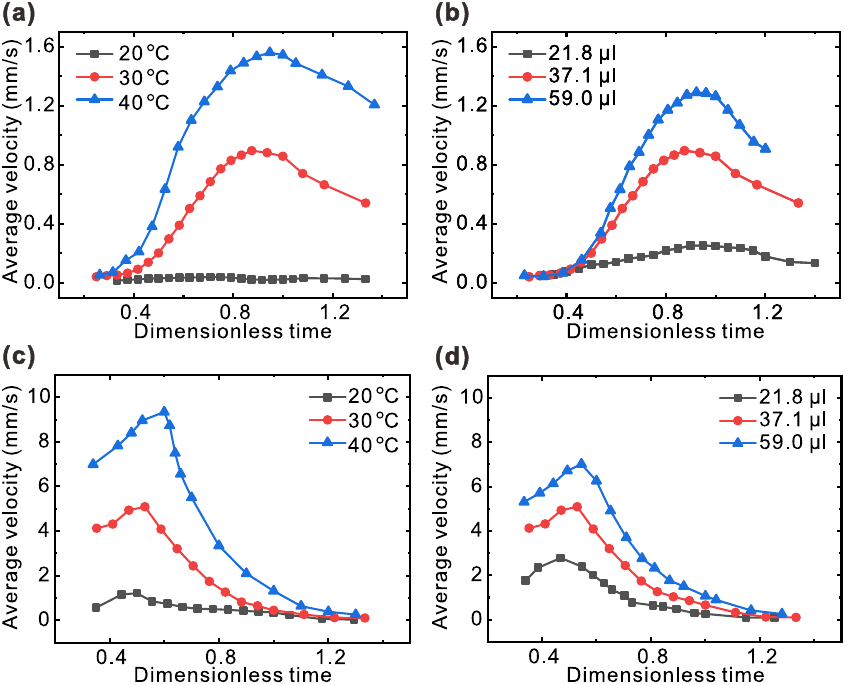}
  \caption{ Average velocity in the melted region of droplets under the influence of heating temperature and droplet volume: (a, c) on the SN superhydrophobic substrate; (b, d) on the HMN superhydrophobic substrate. The dimensionless time is the time normalized by the total melting time of the droplet. The droplet volume is 37.1 $\upmu$l in (a) and (c), and the heating temperature is 30 $^\circ$C in (b) and (d).}\label{fig:06}
\end{figure}

To quantify the time variation of the flow intensity in the droplet, the average velocity magnitude in the melted region is plotted over time in Figure \ref{fig:06}, which shows different trends for the two superhydrophobic substrates during the melting process. On the SN superhydrophobic substrate, the average velocity gradually increases with time and reaches the maximum at the end of ice melting. In contrast, on the HMN superhydrophobic substrate, the average velocity is large in the early stage of melting and decreases with time gradually. For the droplet with 37.1 $\upmu$l at 30 $^\circ$C, the average velocity is 0.20 mm/s at $t^*=0.5$ and 0.86 mm/s at $t^*=1.0$ on the SN superhydrophobic substrate (where $t^*$ is the time normalized by the total melting time of the droplet.), in comparison to 5.01 mm/s at $t^*=0.5$ and 0.45 mm/s at $t^*=1.0$ on the HMN superhydrophobic substrate. Due to the strong optical distortion in the edge region, it is difficult to restore the velocity there accurately (as shown in Figures \ref{fig:05}(b) and \ref{fig:05}(d)). Hence, the average velocity is overestimated on the SN superhydrophobic substrate because of the slow peripheral flow and is underestimated on the HMN superhydrophobic substrate because of the fast edge flow. In general, the melting process on the HMN superhydrophobic substrate has a higher flow intensity than that on the SN superhydrophobic substrate under the same conditions. Besides, the heating temperature and droplet volume can also affect the internal flow of melting droplets. For example, on the SN superhydrophobic substrate, the maximum average melting velocity is 0.039 mm/s at 20 $^\circ$C and has a great increment to 1.56 mm/s at 40 $^\circ$C. As for the influence of droplet volumes, the maximum average melting velocity increases from 0.25 mm/s for 21.8-$\upmu$l droplets to 1.29 mm/s for 59.0-$\upmu$l droplets. The experimental results show that increasing the heating temperature and droplet volume enhances the flow intensity inside the droplet but does not alter the time variation trend of the flow pattern and average velocity on these two superhydrophobic substrates.

\subsection{Mechanism for the melting flow}\label{sec:3.3}
Even though the result of the internal flow during the melting process can explain the difference in droplet melting morphology and melting time, we need to further understand the mechanism of the difference in the melting flow. Then, this section aims to explore the mechanism of the melting flow and the effect of superhydrophobic substrates on the melting flow.

\subsubsection{Marangoni convection and natural convection in melting droplet}\label{sec:3.3.1}
\begin{figure}
  \centering
  \includegraphics[width=0.6\columnwidth]{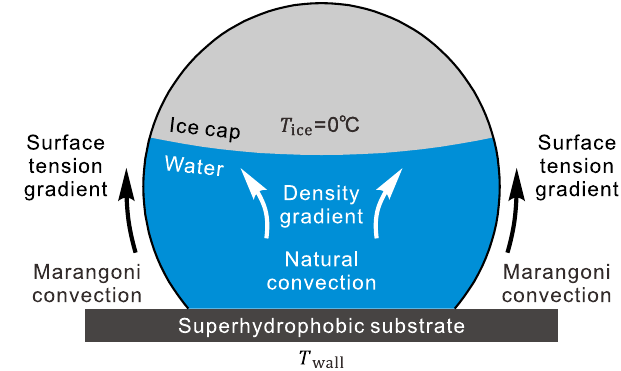}
  \caption{Schematic diagram of Marangoni convection and natural convection in a melting droplet on a superhydrophobic substrate.}\label{fig:07}
\end{figure}

A schematic diagram of droplet melting on a superhydrophobic substrate is shown in Figure \ref{fig:07}. Since the unmelted ice cap always floats in the upper part of the droplet throughout the whole melting process, a relatively stable temperature difference is maintained between the ice cap and the superhydrophobic substrate. The temperature difference induces a temperature gradient along the gas-liquid interface and also inside the droplet. The temperature gradient at the gas-liquid interface leads to a change in the surface tension, which results in the Marangoni convection. The temperature gradient inside the droplet causes a change in the fluid density, which results in the natural convection. According to the temperature gradient direction, the Marangoni convection leads to an upward flow trend along the gas-liquid interface of the droplet, and the natural convection leads to a vertical upward flow trend along the central axis inside the droplet. As the Marangoni convection and natural convection have opposite effects, the specific flow direction depends on the dominant convection mechanism.

Even though Marangoni convection and natural convection were not reported previously for droplet melting problems, they have been reported in many studies of droplet evaporation processes \cite{barmi14, dash14}. The flow in evaporating droplets could be affected by many factors such as temperature field \cite{askounis17, lama20}, droplet volume \cite{alsharafi18}, contact angle \cite{lu11, pan13}, substrate thermal conductivity \cite{chen17}, substrate inclination angle \cite{charitatos21, katre20}, and even particulate matter \cite{parsa18, ren20}, which together determine the flow pattern inside the evaporating droplet. Because of the temperature difference between the unmelted ice cap and the heated substrate, Marangoni convection and natural convection should also exist, which is expected to be more complex due to the melting process.

\subsubsection{Different flow patterns on the two types of substrates}\label{sec:3.3.2}
To estimate the intensity of the Marangoni convection and natural convection, the Marangoni number and the Rayleigh number are often used \cite{chandramohan16}. The Marangoni number characterizes the relative strength of surface tension and viscous force, $Ma={\sigma_T{{\Delta T}}_{Ma}d_{Ma}}/{\left(\alpha\mu\right)}$, where $\sigma_T $ is the temperature coefficient of surface tension variation, ${{\Delta T}}_{Ma}$ is the maximum temperature difference on the droplet surface between the ice cap and the substrate, $d_{Ma}$ is the distance between the ice cap and the substrate along the droplet surface, $\alpha$ is thermal diffusivity, and $\mu $ is the dynamic viscosity of the fluid. The Rayleigh number compares the relative strength of buoyancy and viscous force, and it is defined as $Ra={\beta\rho g{{\Delta T}}_{Ra}d_{Ra}^3}/{(\alpha\mu)}$, where $\beta$ is the volume expansion coefficient due to the temperature change, $\rho$ is the density of the fluid, $g$ is the gravitational acceleration, $d_{Ra}$ is the distance between the ice cap and superhydrophobic substrate along the droplet axis, and ${{\Delta T}}_{Ra}$ is the maximum temperature difference in the droplet axis which has the same value as ${{\Delta T}}_{Ma}$. Due to the constant heating temperature, the maximum temperature difference inside the droplets does not change with time, and the temperature gradient is only related to spatial scales $d_{Ma}$ and $d_{Ra}$. Usually, the two flow patterns coexist, therefore, a ratio of $Ma$ and $Ra$ can be used to indicate the relative strength of the Marangoni convection and natural convection, i.e., ${Ma}/{Ra}={\sigma_T{{\Delta T}}_{Ma}d_{Ma}}/{(\beta\rho g{{\Delta T}}_{Ra}d_{Ra}^3)}$. Hence the ratio ${Ma}/{Ra}$ also indicates the relative strength of surface tension and buoyancy.

\begin{figure}
  \centering
  \includegraphics[width=0.8\columnwidth]{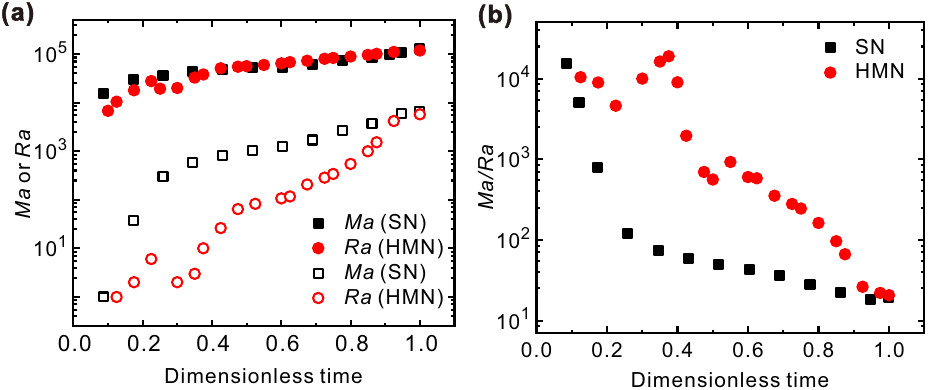}
  \caption{$Ma$ and $Ra$ during the melting process: (a) change of $Ma$ and $Ra$ with time during the melting process on the SN and HMN superhydrophobic substrates; (b) change of $ {Ma}/{Ra}$ with time indicating the relative strength of surface tension and buoyancy. The dimensionless time is the time normalized by the total melting time of the droplet. The heating temperature is 30 $^\circ$C, and the droplet volume is 37.1 $\upmu$l.}\label{fig:08}
\end{figure}

The Marangoni number and the Rayleigh number during a melting process can be calculated by using the dimensions of the droplets based on image processing, and the results for typical melting processes are shown in Figure \ref{fig:08}. Whether on the SN or HMN superhydrophobic substrates, both  $Ma$ and $Ra$ increase gradually with the melting process, indicating that these two forms of convection gradually strengthen. The Marangoni number is always one or two orders of magnitude larger than the Rayleigh number, and ${Ma}/{Ra}$ decreases gradually as the frozen droplet melts. It means that, on the two superhydrophobic substrates, the Marangoni convection is stronger than the natural convection, but the difference in intensity between them gradually decreases.

According to the magnitudes of the dimensionless numbers, $Ma$ and $Ra$, the melting flow pattern should be dominated by the Marangoni convection, and the flow intensity should gradually decrease because of the weakening difference between the Marangoni convection and natural convection during the melting process. This trend is consistent with the droplet melting process on the HMN superhydrophobic substrate but different from that on the SN superhydrophobic substrate. Since there is no critical value to quantify the transition of these two flow trends inside the droplet by comparing the values of $Ma$ and $Ra$, which may be affected by many factors, such as temperature, droplet volume, and contact angle \cite{gelderblom22}, it is difficult to explain the difference in the internal flow pattern and velocity distribution on the SN and HMN superhydrophobic substrates only by this method. Since the experimental conditions of the two melting processes are only different in the structure of superhydrophobic surfaces, the mechanism of the difference in the melting flow is further studied from the perspective of the superhydrophobic surface characteristics.

Superhydrophobic substrates are typically designed according to the method of constructing micro/nano-structures to trap air at the bottom of droplets. The structural strength could be poor and vulnerable to damage, especially under the icing condition \cite{kulinich11, wang13}. During the freezing process, the micro/nano-scale structures of superhydrophobic substrates are partially immersed in the droplet, so they could be destroyed by the volume expansion of ice. We hypothesize that the superhydrophobic particles probably fall off and are preferentially gathered at the gas-liquid interface because of their superhydrophobicity. Therefore, the particles can efficiently restrain the Marangoni convection since this kind of flow is very sensitive to impurities \cite{du22}. If so, the Marangoni convection will be affected, and the flow pattern could be altered.

\begin{figure}
  \centering
  \includegraphics[width=0.7\columnwidth]{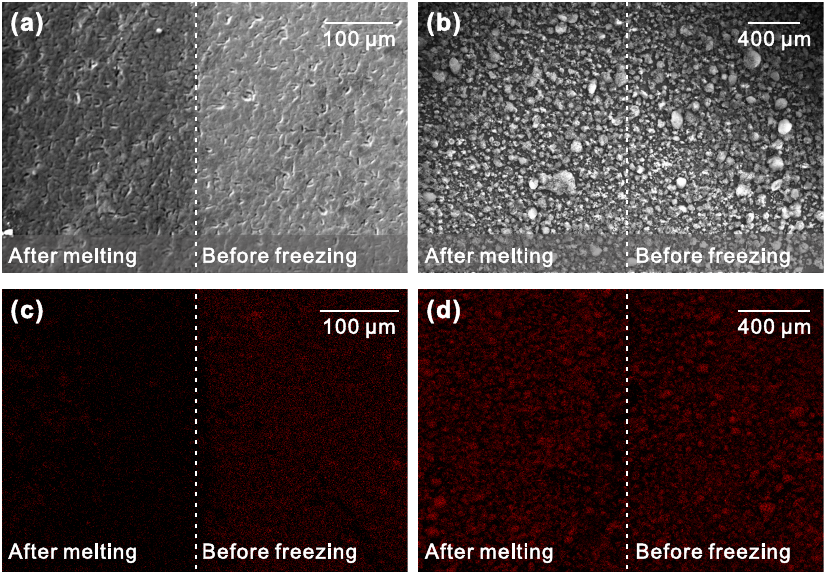}
  \caption{SEM and EDS images of superhydrophobic substrates before freezing and after melting: (a, b) SEM images of the SN (a) and HMN (b) superhydrophobic substrates, showing the morphology change of micro/nano-scale structures, (c, d) EDS images of the SN (c) and HMN (d) superhydrophobic substrates, showing the distribution density of oxygen element (corresponding to silica particles). The right part shows the state before freezing and the left part shows the state after melting. The white dotted line indicates the boundary of the freezing area of the droplets on the substrates.}\label{fig:09}
\end{figure}

To verify the above hypothesis, we compared the micro/nano-scale structures of the substrates before the freezing and after the melting by scanning electron microscopy (SEM). As shown in Figures \ref{fig:09}(a) and \ref{fig:09}(b), the morphology of the SN superhydrophobic substrate surface has more obvious changes after the freezing-melting process, but the HMN superhydrophobic substrate surface is intact and has complete hierarchical micro/nano-scale structures. In addition, we performed energy dispersive spectroscopy (EDS) analyses to check the distribution density of silica particles before the freezing and after the melting on the SN and HMN superhydrophobic substrates. For the SN superhydrophobic substrate, the EDS image indicates that there is an obvious reduction of silica particles after the melting compared with that before the freezing, as shown in Figure \ref{fig:09}(c). In contrast, the EDS image of the HMN superhydrophobic substrate has no obvious difference after the melting and before the freezing, as shown in Figure \ref{fig:09}(d).

\begin{figure}
  \centering
  \includegraphics[width=0.7\columnwidth]{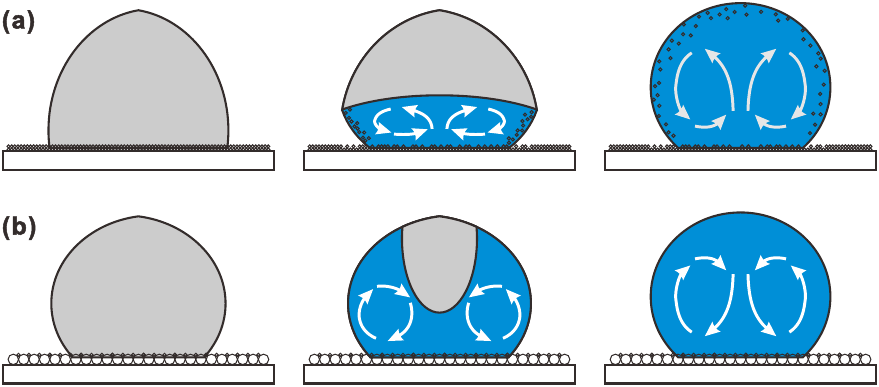}
  \caption{Schematic diagram of droplet before (left), during (middle), and after (right) melting: (a) on the SN superhydrophobic substrate; (b) on the HMN superhydrophobic substrate.}\label{fig:10}
\end{figure}

Based on the above result and considering the presence of both Marangoni convection and natural convection, the flow pattern on the HMN superhydrophobic substrate exhibits a right-counterclockwise and is consistent with the direction dominated by Marangoni convection since the Marangoni number is much larger than the Rayleigh number. As for the SN superhydrophobic substrate, due to the shedding of superhydrophobic particles, the Marangoni convection is suppressed, and the natural convection dominates the melting flow even though the calculated Marangoni number (based on clean gas-liquid interface assumption) is much larger than the Rayleigh number. The whole melting processes on the two superhydrophobic substrates are illustrated in Figure \ref{fig:10}. The most obvious distinction is that the gathering of superhydrophobic particles at the gas-liquid interface leads to the difference in melting morphology and melting flow. Moreover, the average velocity inside the droplet can be used as evidence to further support the mechanism of the different melting processes. For the melting process on the SN superhydrophobic substrate, only the natural convection is considered because of the suppression of Marangoni convection. Its average velocity (see Figure \ref{fig:06}(a)) increases with time, which is consistent with the variation trend of the Rayleigh number (see Figure \ref{fig:08}(a)). While the melting process on the HMN superhydrophobic substrate is the interaction result of Marangoni convection and natural convection, and its average velocity (see Figure \ref{fig:06}(c)) decreases with time except for the early stage, which is consistent with the variation trend of $ {Ma}/{Ra} $ (see Figure \ref{fig:08}(b)).

\subsubsection{Marangoni interfacial flow inhibition by superhydrophobic particles}\label{sec:3.3.3}
To further validate the above analysis, we consider the effect of the superhydrophobic particles on Marangoni interfacial flows. Marangoni convection is very sensitive to impurities including superhydrophobic particles. Superhydrophobic particles can be efficient in suppressing the Marangoni convection since these particles prefer to accumulate at the gas-liquid interface. To observe the effect of particles in a more controlled manner, we designed an extra experiment by measuring the flow in a water pool under a temperature gradient along the horizontal direction. Then, the effect of superhydrophobic particles on Marangoni convection can be studied separately without the influence of phase change.

\begin{figure}
  \centering
  \includegraphics[width=0.45\columnwidth]{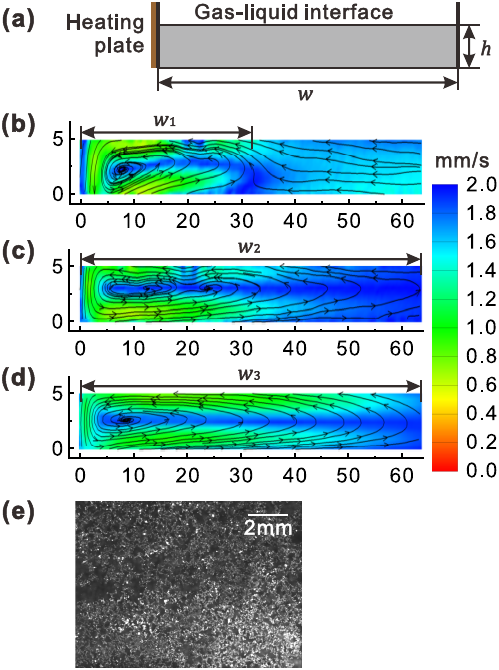}
  \caption{Flow in a water pool under a temperature gradient: (a) Schematic of the experimental setup; (b) Flow field in a water pool with a clean and free gas-liquid interface; (c) Flow field in a water pool with a solid wall on the top; (d) Flow field in a water pool with superhydrophobic particles on the liquid surface; (e) Top-view image showing superhydrophobic particles dispersed on the water surface.}\label{fig:11}
\end{figure}

The schematic of the experimental setup is in Figure \ref{fig:11}(a). A water pool with a large width-to-height ratio is used ($h = 5$ mm and $w = 100$ mm), hence the effect of the right wall can be minimized \cite{schatz01}. The electric heating paste is used on the left side of the pool to provide a fixed high temperature of 40 $^\circ$C. The unheated region of the water pool is maintained at 23--25 $^\circ$C, which is the same as the room temperature in the laboratory. The fluorescent particles used in the PIV measurement of the flow in the water pool are the same as in the experiment in Figure \ref{fig:05}. Three types of boundary conditions were applied on the top of the liquid surface, namely a free liquid surface, a solid plate, and a liquid surface with uniformly dispersed particles. The particles used here are superhydrophobic silica powder with an average diameter of $50 \pm 5$ nm, which only floats at the gas-liquid interface and are difficult to enter the interior of the water pool (as shown in Figure \ref{fig:11}(e)). For the free liquid surface boundary condition, under the temperature gradient in the horizontal direction, the Marangoni convection and natural convection are established with many vortices along the direction of the gas-liquid interface \cite{mizev09}, and the first vortex is shown in Figure \ref{fig:11}(b). When superhydrophobic particles are uniformly dispersed on the water surface (see the top view in Figure \ref{fig:11}(e)), the flow under the temperature gradient (see Figure \ref{fig:11}(d)) is the same as that of replacing the gas-liquid interface with a solid wall (see Figure \ref{fig:11}(c)), both of which have only a big main vortex. Comparing the width of the first vortex from the left wall, we can see that $w_2$ in Figure \ref{fig:11}(c) and $w_3$ in Figure \ref{fig:11}(d) are much larger than $w_1$ in Figure \ref{fig:11}(b). It is easy to know that the surface tension gradient is eliminated in Figure \ref{fig:11}(c) because there is no free surface, so superhydrophobic particles in Figure \ref{fig:11}(d) have a similar effect as the solid wall in Figure \ref{fig:11}(c). In this experiment, dispersed superhydrophobic particles eliminate the effect of surface tension gradient and inhibit the Marangoni convection. Therefore, it demonstrates that the superhydrophobic particles shed from the SN superhydrophobic substrate in the freezing/melting cycle have the same effect of inhibiting the Marangoni convection during the melting process.

\section{Conclusions}\label{sec:4}
In this study, the melting process of single frozen droplets under active heating is investigated and the effect of the superhydrophobic surface on the melting flow is considered. The results show a remarkable difference in the melting morphology and melting time for droplets on the SN and HMN superhydrophobic substrates. On the HMN superhydrophobic substrate, the frozen droplet preferentially melts on both sides and the melting flow direction is the same as Marangoni convection with the counterclockwise flow on the right part. While on the SN superhydrophobic substrates, the frozen droplet preferentially melts in the middle and the melting flow direction is the same as the natural convection with the clockwise flow on the right part. In addition, the melting time on the HMN superhydrophobic substrate is shorter than that on the SN superhydrophobic substrate, showing a superior deicing performance.

The difference in the melting morphology and melting time for droplets on the SN and HMN superhydrophobic substrates could be attributed to the internal flow during the melting process. The temperature gradient in the droplet can result in Marangoni convection and natural convection. For the HMN superhydrophobic substrate, the internal flow is dominated by the Marangoni convection due to the temperature gradient along the droplet surface; while for the SN superhydrophobic substrate, the Marangoni convection is inhibited by the superhydrophobic particles at the droplet surface shed from the fragile superhydrophobic substrate during the freezing-melting process.

%
%
\begin{acknowledgement}
This work is supported by the National Natural Science Foundation of China (Grant Nos.\ 52176083, 51920105010, and 51921004).
\end{acknowledgement}
%
%
\begin{suppinfo}

The following files are available free of charge.
\begin{itemize}
  \item Movie 1: Video clips showing the melting process on SN superhydrophobic surfaces, corresponding to Figure \ref{fig:03}a.
  \item Movie 2: Video clip showing the melting process on HMN superhydrophobic surfaces, corresponding to Figure \ref{fig:03}c.
  \item PDF file: Melting process on other superhydrophobic surfaces.
\end{itemize}

\end{suppinfo}


\bibliography{MeltingDroplet}

\providecommand{\latin}[1]{#1}
\makeatletter
\providecommand{\doi}
  {\begingroup\let\do\@makeother\dospecials
  \catcode`\{=1 \catcode`\}=2 \doi@aux}
\providecommand{\doi@aux}[1]{\endgroup\texttt{#1}}
\makeatother
\providecommand*\mcitethebibliography{\thebibliography}
\csname @ifundefined\endcsname{endmcitethebibliography}
  {\let\endmcitethebibliography\endthebibliography}{}
\begin{mcitethebibliography}{53}
\providecommand*\natexlab[1]{#1}
\providecommand*\mciteSetBstSublistMode[1]{}
\providecommand*\mciteSetBstMaxWidthForm[2]{}
\providecommand*\mciteBstWouldAddEndPuncttrue
  {\def\EndOfBibitem{\unskip.}}
\providecommand*\mciteBstWouldAddEndPunctfalse
  {\let\EndOfBibitem\relax}
\providecommand*\mciteSetBstMidEndSepPunct[3]{}
\providecommand*\mciteSetBstSublistLabelBeginEnd[3]{}
\providecommand*\EndOfBibitem{}
\mciteSetBstSublistMode{f}
\mciteSetBstMaxWidthForm{subitem}{(\alph{mcitesubitemcount})}
\mciteSetBstSublistLabelBeginEnd
  {\mcitemaxwidthsubitemform\space}
  {\relax}
  {\relax}

\bibitem[Yamazaki \latin{et~al.}(2021)Yamazaki, Jemcov, and Sakaue]{yamazaki21}
Yamazaki,~M.; Jemcov,~A.; Sakaue,~H. A review on the current status of icing
  physics and mitigation in aviation. \emph{Aerospace} \textbf{2021}, \emph{8},
  188\relax
\mciteBstWouldAddEndPuncttrue
\mciteSetBstMidEndSepPunct{\mcitedefaultmidpunct}
{\mcitedefaultendpunct}{\mcitedefaultseppunct}\relax
\EndOfBibitem
\bibitem[Luo and Jiao(2018)Luo, and Jiao]{luo18}
Luo,~Y.; Jiao,~K. Cold start of proton exchange membrane fuel cell. \emph{Prog.
  Energy Combust. Sci.} \textbf{2018}, \emph{64}, 29--61\relax
\mciteBstWouldAddEndPuncttrue
\mciteSetBstMidEndSepPunct{\mcitedefaultmidpunct}
{\mcitedefaultendpunct}{\mcitedefaultseppunct}\relax
\EndOfBibitem
\bibitem[Wei \latin{et~al.}(2020)Wei, Yang, Zuo, and Zhong]{wei19}
Wei,~K.; Yang,~Y.; Zuo,~H.; Zhong,~D. A review on ice detection technology and
  ice elimination technology for wind turbine. \emph{Wind Energy}
  \textbf{2020}, \emph{23}, 433--457\relax
\mciteBstWouldAddEndPuncttrue
\mciteSetBstMidEndSepPunct{\mcitedefaultmidpunct}
{\mcitedefaultendpunct}{\mcitedefaultseppunct}\relax
\EndOfBibitem
\bibitem[Boreyko and Chen(2009)Boreyko, and Chen]{boreyko09}
Boreyko,~J.~B.; Chen,~C.~H. Self-propelled dropwise condensate on
  superhydrophobic surfaces. \emph{Phys. Rev. Lett.} \textbf{2009}, \emph{103},
  184501\relax
\mciteBstWouldAddEndPuncttrue
\mciteSetBstMidEndSepPunct{\mcitedefaultmidpunct}
{\mcitedefaultendpunct}{\mcitedefaultseppunct}\relax
\EndOfBibitem
\bibitem[Feng \latin{et~al.}(2012)Feng, Qin, and Yao]{feng12-frAQB}
Feng,~J.; Qin,~Z.; Yao,~S. Factors affecting the spontaneous motion of
  condensate drops on superhydrophobic copper surfaces. \emph{Langmuir}
  \textbf{2012}, \emph{28}, 6067--6075\relax
\mciteBstWouldAddEndPuncttrue
\mciteSetBstMidEndSepPunct{\mcitedefaultmidpunct}
{\mcitedefaultendpunct}{\mcitedefaultseppunct}\relax
\EndOfBibitem
\bibitem[Liu \latin{et~al.}(2021)Liu, Wang, Zhang, and Chen]{liu21DropletJump}
Liu,~X.; Wang,~P.; Zhang,~D.; Chen,~X. Atmospheric corrosion protection
  performance and mechanism of superhydrophobic surface based on
  coalescence-induced droplet self-jumping behavior. \emph{ACS Appl. Mater.
  Interfaces} \textbf{2021}, \emph{13}, 25438--25450\relax
\mciteBstWouldAddEndPuncttrue
\mciteSetBstMidEndSepPunct{\mcitedefaultmidpunct}
{\mcitedefaultendpunct}{\mcitedefaultseppunct}\relax
\EndOfBibitem
\bibitem[Wen \latin{et~al.}(2014)Wen, Wang, Zhang, Jiang, and Zheng]{wen14}
Wen,~M.; Wang,~L.; Zhang,~M.; Jiang,~L.; Zheng,~Y. Antifogging and icing-delay
  properties of composite micro- and nanostructured surfaces. \emph{ACS Appl.
  Mater. Interfaces} \textbf{2014}, \emph{6}, 3963--3968\relax
\mciteBstWouldAddEndPuncttrue
\mciteSetBstMidEndSepPunct{\mcitedefaultmidpunct}
{\mcitedefaultendpunct}{\mcitedefaultseppunct}\relax
\EndOfBibitem
\bibitem[Wen \latin{et~al.}(2017)Wen, Xu, Zhao, Lee, Ma, and Yang]{wen17}
Wen,~R.; Xu,~S.; Zhao,~D.; Lee,~Y.~C.; Ma,~X.; Yang,~R. Hierarchical
  superhydrophobic surfaces with micropatterned nanowire arrays for
  high-efficiency jumping droplet condensation. \emph{ACS Appl. Mater.
  Interfaces} \textbf{2017}, \emph{9}, 44911--44921\relax
\mciteBstWouldAddEndPuncttrue
\mciteSetBstMidEndSepPunct{\mcitedefaultmidpunct}
{\mcitedefaultendpunct}{\mcitedefaultseppunct}\relax
\EndOfBibitem
\bibitem[Chu \latin{et~al.}(2017)Chu, Wu, and Ma]{chu17}
Chu,~F.; Wu,~X.; Ma,~Q. Condensed droplet growth on surfaces with various
  wettability. \emph{Appl. Therm. Eng.} \textbf{2017}, \emph{115},
  1101--1108\relax
\mciteBstWouldAddEndPuncttrue
\mciteSetBstMidEndSepPunct{\mcitedefaultmidpunct}
{\mcitedefaultendpunct}{\mcitedefaultseppunct}\relax
\EndOfBibitem
\bibitem[Alizadeh \latin{et~al.}(2012)Alizadeh, Yamada, Li, Shang, Otta, Zhong,
  Ge, Dhinojwala, Conway, Bahadur, Vinciquerra, Stephens, and
  Blohm]{alizadeh12}
Alizadeh,~A.; Yamada,~M.; Li,~R.; Shang,~W.; Otta,~S.; Zhong,~S.; Ge,~L.;
  Dhinojwala,~A.; Conway,~K.~R.; Bahadur,~V.; Vinciquerra,~A.~J.; Stephens,~B.;
  Blohm,~M.~L. Dynamics of ice nucleation on water repellent surfaces.
  \emph{Langmuir} \textbf{2012}, \emph{28}, 3180--3186\relax
\mciteBstWouldAddEndPuncttrue
\mciteSetBstMidEndSepPunct{\mcitedefaultmidpunct}
{\mcitedefaultendpunct}{\mcitedefaultseppunct}\relax
\EndOfBibitem
\bibitem[Jung \latin{et~al.}(2011)Jung, Dorrestijn, Raps, Das, Megaridis, and
  Poulikakos]{jung11}
Jung,~S.; Dorrestijn,~M.; Raps,~D.; Das,~A.; Megaridis,~C.~M.; Poulikakos,~D.
  Are superhydrophobic surfaces best for icephobicity? \emph{Langmuir}
  \textbf{2011}, \emph{27}, 3059--3066\relax
\mciteBstWouldAddEndPuncttrue
\mciteSetBstMidEndSepPunct{\mcitedefaultmidpunct}
{\mcitedefaultendpunct}{\mcitedefaultseppunct}\relax
\EndOfBibitem
\bibitem[Nosonovsky and Hejazi(2012)Nosonovsky, and Hejazi]{nosonovsky12}
Nosonovsky,~M.; Hejazi,~V. Why superhydrophobic surfaces are not always
  icephobic. \emph{ACS Nano} \textbf{2012}, \emph{6}, 8488–8491\relax
\mciteBstWouldAddEndPuncttrue
\mciteSetBstMidEndSepPunct{\mcitedefaultmidpunct}
{\mcitedefaultendpunct}{\mcitedefaultseppunct}\relax
\EndOfBibitem
\bibitem[Montes Ruiz~Cabello \latin{et~al.}(2021)Montes Ruiz~Cabello,
  Bermudez~Romero, Ibanez~Ibanez, Cabrerizo~Vílchez, and
  Rodríguez~Valverde]{montes21}
Montes Ruiz~Cabello,~F.~J.; Bermudez~Romero,~S.; Ibanez~Ibanez,~P.~F.;
  Cabrerizo~Vílchez,~M.~A.; Rodríguez~Valverde,~M.~A. Freezing delay of
  sessile drops: Probing the impact of contact angle, surface roughness and
  thermal conductivity. \emph{Appl. Surf. Sci.} \textbf{2021}, \emph{537},
  147964\relax
\mciteBstWouldAddEndPuncttrue
\mciteSetBstMidEndSepPunct{\mcitedefaultmidpunct}
{\mcitedefaultendpunct}{\mcitedefaultseppunct}\relax
\EndOfBibitem
\bibitem[Shi and Duan(2022)Shi, and Duan]{shi22}
Shi,~K.; Duan,~X. Freezing delay of water droplets on metallic hydrophobic
  surfaces in a cold environment. \emph{Appl. Therm. Eng.} \textbf{2022},
  \emph{216}, 119131\relax
\mciteBstWouldAddEndPuncttrue
\mciteSetBstMidEndSepPunct{\mcitedefaultmidpunct}
{\mcitedefaultendpunct}{\mcitedefaultseppunct}\relax
\EndOfBibitem
\bibitem[Zhang \latin{et~al.}(2023)Zhang, Song, Chao, and Shen]{zhang23}
Zhang,~L.; Song,~M.; Chao,~C. Y.~H.; Shen,~J. An experimental study on the
  dynamic frosting characteristics on the edge zone of a horizontal copper
  plate under forced convection. \emph{Int. J. Heat Mass Transfer}
  \textbf{2023}, \emph{200}, 123541\relax
\mciteBstWouldAddEndPuncttrue
\mciteSetBstMidEndSepPunct{\mcitedefaultmidpunct}
{\mcitedefaultendpunct}{\mcitedefaultseppunct}\relax
\EndOfBibitem
\bibitem[Stamatopoulos \latin{et~al.}(2017)Stamatopoulos, Hemrle, Wang, and
  Poulikakos]{stamatopoulos17}
Stamatopoulos,~C.; Hemrle,~J.; Wang,~D.; Poulikakos,~D. Exceptional anti-icing
  performance of self-impregnating slippery surfaces. \emph{ACS Appl. Mater.
  Interfaces} \textbf{2017}, \emph{9}, 10233--10242\relax
\mciteBstWouldAddEndPuncttrue
\mciteSetBstMidEndSepPunct{\mcitedefaultmidpunct}
{\mcitedefaultendpunct}{\mcitedefaultseppunct}\relax
\EndOfBibitem
\bibitem[Wang \latin{et~al.}(2013)Wang, Xue, Wang, Chen, and Ding]{wang13}
Wang,~Y.; Xue,~J.; Wang,~Q.; Chen,~Q.; Ding,~J. Verification of
  icephobic/anti-icing properties of a superhydrophobic surface. \emph{ACS
  Appl. Mater. Interfaces} \textbf{2013}, \emph{5}, 3370--81\relax
\mciteBstWouldAddEndPuncttrue
\mciteSetBstMidEndSepPunct{\mcitedefaultmidpunct}
{\mcitedefaultendpunct}{\mcitedefaultseppunct}\relax
\EndOfBibitem
\bibitem[Bharathidasan \latin{et~al.}(2014)Bharathidasan, Kumar, Bobji,
  Chakradhar, and Basu]{bharathidasan14}
Bharathidasan,~T.; Kumar,~S.~V.; Bobji,~M.~S.; Chakradhar,~R. P.~S.;
  Basu,~B.~J. Effect of wettability and surface roughness on ice-adhesion
  strength of hydrophilic, hydrophobic and superhydrophobic surfaces.
  \emph{Appl. Surf. Sci.} \textbf{2014}, \emph{314}, 241--250\relax
\mciteBstWouldAddEndPuncttrue
\mciteSetBstMidEndSepPunct{\mcitedefaultmidpunct}
{\mcitedefaultendpunct}{\mcitedefaultseppunct}\relax
\EndOfBibitem
\bibitem[Chen \latin{et~al.}(2012)Chen, Liu, He, Li, Cui, Zhang, Zeng, Zhang,
  Wang, and Song]{chen12}
Chen,~J.; Liu,~J.; He,~M.; Li,~K.; Cui,~D.; Zhang,~Q.; Zeng,~X.; Zhang,~Y.;
  Wang,~J.; Song,~Y. Superhydrophobic surfaces cannot reduce ice adhesion.
  \emph{Appl. Phys. Lett.} \textbf{2012}, \emph{101}, 111603\relax
\mciteBstWouldAddEndPuncttrue
\mciteSetBstMidEndSepPunct{\mcitedefaultmidpunct}
{\mcitedefaultendpunct}{\mcitedefaultseppunct}\relax
\EndOfBibitem
\bibitem[Davis \latin{et~al.}(2014)Davis, Yeong, Steele, Bayer, and
  Loth]{davis14}
Davis,~A.; Yeong,~Y.~H.; Steele,~A.; Bayer,~I.~S.; Loth,~E. Superhydrophobic
  nanocomposite surface topography and ice adhesion. \emph{ACS Appl. Mater.
  Interfaces} \textbf{2014}, \emph{6}, 9272--9\relax
\mciteBstWouldAddEndPuncttrue
\mciteSetBstMidEndSepPunct{\mcitedefaultmidpunct}
{\mcitedefaultendpunct}{\mcitedefaultseppunct}\relax
\EndOfBibitem
\bibitem[Murphy \latin{et~al.}(2017)Murphy, McClintic, Lester, Collier, and
  Boreyko]{murphy17}
Murphy,~K.~R.; McClintic,~W.~T.; Lester,~K.~C.; Collier,~C.~P.; Boreyko,~J.~B.
  Dynamic Defrosting on Scalable Superhydrophobic Surfaces. \emph{ACS Appl.
  Mater. Interfaces} \textbf{2017}, \emph{9}, 24308--24317\relax
\mciteBstWouldAddEndPuncttrue
\mciteSetBstMidEndSepPunct{\mcitedefaultmidpunct}
{\mcitedefaultendpunct}{\mcitedefaultseppunct}\relax
\EndOfBibitem
\bibitem[Boreyko \latin{et~al.}(2013)Boreyko, Srijanto, Nguyen, Vega,
  Fuentes-Cabrera, and Collier]{boreyko13}
Boreyko,~J.~B.; Srijanto,~B.~R.; Nguyen,~T.~D.; Vega,~C.; Fuentes-Cabrera,~M.;
  Collier,~C.~P. Dynamic defrosting on nanostructured superhydrophobic
  surfaces. \emph{Langmuir} \textbf{2013}, \emph{29}, 9516--9524\relax
\mciteBstWouldAddEndPuncttrue
\mciteSetBstMidEndSepPunct{\mcitedefaultmidpunct}
{\mcitedefaultendpunct}{\mcitedefaultseppunct}\relax
\EndOfBibitem
\bibitem[Chen \latin{et~al.}(2021)Chen, Meng, Liu, Li, and Miao]{chen21}
Chen,~A.; Meng,~Y.; Liu,~B.; Li,~Y.; Miao,~Z. Effects of inclination on the
  frosting process on cold surface of copper heat exchanger. \emph{Energy
  Build.} \textbf{2021}, \emph{231}, 110628\relax
\mciteBstWouldAddEndPuncttrue
\mciteSetBstMidEndSepPunct{\mcitedefaultmidpunct}
{\mcitedefaultendpunct}{\mcitedefaultseppunct}\relax
\EndOfBibitem
\bibitem[Zhu \latin{et~al.}(2017)Zhu, Liu, Shen, Tao, Wang, and Pan]{zhu17}
Zhu,~C.; Liu,~S.; Shen,~Y.; Tao,~J.; Wang,~G.; Pan,~L. Verifying the deicing
  capacity of superhydrophobic anti-icing surfaces based on wind and thermal
  fields. \emph{Surf. Coat. Technol.} \textbf{2017}, \emph{309}, 703--708\relax
\mciteBstWouldAddEndPuncttrue
\mciteSetBstMidEndSepPunct{\mcitedefaultmidpunct}
{\mcitedefaultendpunct}{\mcitedefaultseppunct}\relax
\EndOfBibitem
\bibitem[Huang \latin{et~al.}(2020)Huang, Wang, Zhang, Xu, and Li]{huang20}
Huang,~Z.; Wang,~F.; Zhang,~R.; Xu,~W.; Li,~J. Self-ejections of multiple
  isolated slushes on disorderly grooved superhydrophobic surfaces. \emph{Appl.
  Phys. Lett.} \textbf{2020}, \emph{116}, 053702\relax
\mciteBstWouldAddEndPuncttrue
\mciteSetBstMidEndSepPunct{\mcitedefaultmidpunct}
{\mcitedefaultendpunct}{\mcitedefaultseppunct}\relax
\EndOfBibitem
\bibitem[Liu \latin{et~al.}(2021)Liu, Chen, Zhao, Zhu, Wang, Chen, and
  Zhang]{liu21}
Liu,~X.; Chen,~H.; Zhao,~Z.; Zhu,~Y.; Wang,~Z.; Chen,~J.; Zhang,~D. Tunable
  self-jumping of melting frost on macro-patterned anisotropic superhydrophobic
  surfaces. \emph{Surf. Coat. Technol.} \textbf{2021}, \emph{409}, 126858\relax
\mciteBstWouldAddEndPuncttrue
\mciteSetBstMidEndSepPunct{\mcitedefaultmidpunct}
{\mcitedefaultendpunct}{\mcitedefaultseppunct}\relax
\EndOfBibitem
\bibitem[Tang(2023)]{tang2023new1}
Tang,~X. Multifunctional droplet-surface interaction effected by bulk
  properties. \emph{Droplet} \textbf{2023}, \emph{2}, e38\relax
\mciteBstWouldAddEndPuncttrue
\mciteSetBstMidEndSepPunct{\mcitedefaultmidpunct}
{\mcitedefaultendpunct}{\mcitedefaultseppunct}\relax
\EndOfBibitem
\bibitem[Lathia \latin{et~al.}(2023)Lathia, Modak, and Sen]{lathia2023new2}
Lathia,~R.; Modak,~C.~D.; Sen,~P. Two modes of contact-time reduction in the
  impact of particle-coated droplets on superhydrophobic surfaces.
  \emph{Droplet} \textbf{2023}, e89\relax
\mciteBstWouldAddEndPuncttrue
\mciteSetBstMidEndSepPunct{\mcitedefaultmidpunct}
{\mcitedefaultendpunct}{\mcitedefaultseppunct}\relax
\EndOfBibitem
\bibitem[Li \latin{et~al.}(2022)Li, Li, Dang, and Liu]{li22}
Li,~Y.; Li,~M.; Dang,~C.; Liu,~X. Effects of dissolved gas on the nucleation
  and growth of ice crystals in freezing droplets. \emph{Int. J. Heat Mass
  Transfer} \textbf{2022}, \emph{184}, 122334\relax
\mciteBstWouldAddEndPuncttrue
\mciteSetBstMidEndSepPunct{\mcitedefaultmidpunct}
{\mcitedefaultendpunct}{\mcitedefaultseppunct}\relax
\EndOfBibitem
\bibitem[Chu \latin{et~al.}(2019)Chu, Zhang, Li, Jin, Zhang, Wu, and
  Wen]{Chu2019new3}
Chu,~F.; Zhang,~X.; Li,~S.; Jin,~H.; Zhang,~J.; Wu,~X.; Wen,~D. Bubble
  formation in freezing droplets. \emph{Phys. Rev. Fluids} \textbf{2019},
  \emph{4}, 071601\relax
\mciteBstWouldAddEndPuncttrue
\mciteSetBstMidEndSepPunct{\mcitedefaultmidpunct}
{\mcitedefaultendpunct}{\mcitedefaultseppunct}\relax
\EndOfBibitem
\bibitem[Wang \latin{et~al.}(2022)Wang, Tian, Jiang, Luo, Chen, Hu, Zhang, and
  Zhong]{wang22}
Wang,~L.; Tian,~Z.; Jiang,~G.; Luo,~X.; Chen,~C.; Hu,~X.; Zhang,~H.; Zhong,~M.
  Spontaneous dewetting transitions of droplets during icing \& melting cycle.
  \emph{Nat. Commun.} \textbf{2022}, \emph{13}, 378\relax
\mciteBstWouldAddEndPuncttrue
\mciteSetBstMidEndSepPunct{\mcitedefaultmidpunct}
{\mcitedefaultendpunct}{\mcitedefaultseppunct}\relax
\EndOfBibitem
\bibitem[Feng \latin{et~al.}(2012)Feng, Pang, Qin, Ma, and Yao]{feng12}
Feng,~J.; Pang,~Y.; Qin,~Z.; Ma,~R.; Yao,~S. Why condensate drops can
  spontaneously move away on some superhydrophobic surfaces but not on others.
  \emph{ACS Appl. Mater. Interfaces} \textbf{2012}, \emph{4}, 6618--6625\relax
\mciteBstWouldAddEndPuncttrue
\mciteSetBstMidEndSepPunct{\mcitedefaultmidpunct}
{\mcitedefaultendpunct}{\mcitedefaultseppunct}\relax
\EndOfBibitem
\bibitem[Chu \latin{et~al.}(2019)Chu, Gao, Zhang, Wu, and Wen]{chu2019new4}
Chu,~F.; Gao,~S.; Zhang,~X.; Wu,~X.; Wen,~D. {Droplet re-icing characteristics
  on a superhydrophobic surface}. \emph{Appl. Phys. Lett.} \textbf{2019},
  \emph{115}, 073703\relax
\mciteBstWouldAddEndPuncttrue
\mciteSetBstMidEndSepPunct{\mcitedefaultmidpunct}
{\mcitedefaultendpunct}{\mcitedefaultseppunct}\relax
\EndOfBibitem
\bibitem[Kang \latin{et~al.}(2004)Kang, Lee, Lee, and Kang]{kang04}
Kang,~K.~H.; Lee,~S.~J.; Lee,~C.~M.; Kang,~I.~S. Quantitative visualization of
  flow inside an evaporating droplet using the ray tracing method. \emph{Meas.
  Sci. Technol.} \textbf{2004}, \emph{15}, 1104--1112\relax
\mciteBstWouldAddEndPuncttrue
\mciteSetBstMidEndSepPunct{\mcitedefaultmidpunct}
{\mcitedefaultendpunct}{\mcitedefaultseppunct}\relax
\EndOfBibitem
\bibitem[Barmi and Meinhart(2014)Barmi, and Meinhart]{barmi14}
Barmi,~M.~R.; Meinhart,~C.~D. Convective flows in evaporating sessile droplets.
  \emph{J. Phys. Chem. B} \textbf{2014}, \emph{118}, 2414--2421\relax
\mciteBstWouldAddEndPuncttrue
\mciteSetBstMidEndSepPunct{\mcitedefaultmidpunct}
{\mcitedefaultendpunct}{\mcitedefaultseppunct}\relax
\EndOfBibitem
\bibitem[Dash \latin{et~al.}(2014)Dash, Chandramohan, Weibel, and
  Garimella]{dash14}
Dash,~S.; Chandramohan,~A.; Weibel,~J.~A.; Garimella,~S.~V. Buoyancy-induced
  on-the-spot mixing in droplets evaporating on nonwetting surfaces.
  \emph{Phys. Rev. E} \textbf{2014}, \emph{90}, 062407\relax
\mciteBstWouldAddEndPuncttrue
\mciteSetBstMidEndSepPunct{\mcitedefaultmidpunct}
{\mcitedefaultendpunct}{\mcitedefaultseppunct}\relax
\EndOfBibitem
\bibitem[Askounis \latin{et~al.}(2017)Askounis, Kita, Kohno, Takata, Koutsos,
  and Sefiane]{askounis17}
Askounis,~A.; Kita,~Y.; Kohno,~M.; Takata,~Y.; Koutsos,~V.; Sefiane,~K.
  Influence of local heating on Marangoni flows and evaporation kinetics of
  pure water drops. \emph{Langmuir} \textbf{2017}, \emph{33}, 5666--5674\relax
\mciteBstWouldAddEndPuncttrue
\mciteSetBstMidEndSepPunct{\mcitedefaultmidpunct}
{\mcitedefaultendpunct}{\mcitedefaultseppunct}\relax
\EndOfBibitem
\bibitem[Lama \latin{et~al.}(2020)Lama, Satapathy, and Basavaraj]{lama20}
Lama,~H.; Satapathy,~D.~K.; Basavaraj,~M.~G. Modulation of central depletion
  zone in evaporated sessile drops via substrate heating. \emph{Langmuir}
  \textbf{2020}, \emph{36}, 4737--4744\relax
\mciteBstWouldAddEndPuncttrue
\mciteSetBstMidEndSepPunct{\mcitedefaultmidpunct}
{\mcitedefaultendpunct}{\mcitedefaultseppunct}\relax
\EndOfBibitem
\bibitem[Al-Sharafi \latin{et~al.}(2018)Al-Sharafi, Yilbas, and
  Al-Zahrani]{alsharafi18}
Al-Sharafi,~A.; Yilbas,~B.~S.; Al-Zahrani,~A. Ferro-liquid droplet heat
  transfer on water surface: effect of droplet volume on droplet fluidity.
  \emph{J. Thermophys. Heat Transfer} \textbf{2018}, \emph{32},
  1072--1087\relax
\mciteBstWouldAddEndPuncttrue
\mciteSetBstMidEndSepPunct{\mcitedefaultmidpunct}
{\mcitedefaultendpunct}{\mcitedefaultseppunct}\relax
\EndOfBibitem
\bibitem[Lu \latin{et~al.}(2011)Lu, Duan, Wang, and Lee]{lu11}
Lu,~G.; Duan,~Y.-Y.; Wang,~X.-D.; Lee,~D.-J. Internal flow in evaporating
  droplet on heated solid surface. \emph{Int. J. Heat Mass Transfer}
  \textbf{2011}, \emph{54}, 4437--4447\relax
\mciteBstWouldAddEndPuncttrue
\mciteSetBstMidEndSepPunct{\mcitedefaultmidpunct}
{\mcitedefaultendpunct}{\mcitedefaultseppunct}\relax
\EndOfBibitem
\bibitem[Pan \latin{et~al.}(2013)Pan, Dash, Weibel, and Garimella]{pan13}
Pan,~Z.; Dash,~S.; Weibel,~J.~A.; Garimella,~S.~V. Assessment of water droplet
  evaporation mechanisms on hydrophobic and superhydrophobic substrates.
  \emph{Langmuir} \textbf{2013}, \emph{29}, 15831--15841\relax
\mciteBstWouldAddEndPuncttrue
\mciteSetBstMidEndSepPunct{\mcitedefaultmidpunct}
{\mcitedefaultendpunct}{\mcitedefaultseppunct}\relax
\EndOfBibitem
\bibitem[Chen \latin{et~al.}(2017)Chen, Hu, Wang, Hong, and Cheng]{chen17}
Chen,~Y.~H.; Hu,~W.~N.; Wang,~J.; Hong,~F.~J.; Cheng,~P. Transient effects and
  mass convection in sessile droplet evaporation: The role of liquid and
  substrate thermophysical properties. \emph{Int. J. Heat Mass Transfer}
  \textbf{2017}, \emph{108}, 2072--2087\relax
\mciteBstWouldAddEndPuncttrue
\mciteSetBstMidEndSepPunct{\mcitedefaultmidpunct}
{\mcitedefaultendpunct}{\mcitedefaultseppunct}\relax
\EndOfBibitem
\bibitem[Charitatos \latin{et~al.}(2021)Charitatos, Pham, and
  Kumar]{charitatos21}
Charitatos,~V.; Pham,~T.; Kumar,~S. Droplet evaporation on inclined substrates.
  \emph{Phys. Rev. Fluids} \textbf{2021}, \emph{6}, 084001\relax
\mciteBstWouldAddEndPuncttrue
\mciteSetBstMidEndSepPunct{\mcitedefaultmidpunct}
{\mcitedefaultendpunct}{\mcitedefaultseppunct}\relax
\EndOfBibitem
\bibitem[Katre \latin{et~al.}(2020)Katre, Gurrala, Balusamy, Banerjee, and
  Sahu]{katre20}
Katre,~P.; Gurrala,~P.; Balusamy,~S.; Banerjee,~S.; Sahu,~K.~C. Evaporation of
  sessile ethanol-water droplets on a critically inclined heated surface.
  \emph{Int. J. Multiphase Flow} \textbf{2020}, \emph{131}, 103368\relax
\mciteBstWouldAddEndPuncttrue
\mciteSetBstMidEndSepPunct{\mcitedefaultmidpunct}
{\mcitedefaultendpunct}{\mcitedefaultseppunct}\relax
\EndOfBibitem
\bibitem[Parsa \latin{et~al.}(2018)Parsa, Harmand, and Sefiane]{parsa18}
Parsa,~M.; Harmand,~S.; Sefiane,~K. Mechanisms of pattern formation from dried
  sessile drops. \emph{Adv. Colloid Interface Sci.} \textbf{2018}, \emph{254},
  22--47\relax
\mciteBstWouldAddEndPuncttrue
\mciteSetBstMidEndSepPunct{\mcitedefaultmidpunct}
{\mcitedefaultendpunct}{\mcitedefaultseppunct}\relax
\EndOfBibitem
\bibitem[Ren \latin{et~al.}(2020)Ren, Crivoi, and Duan]{ren20}
Ren,~J.; Crivoi,~A.; Duan,~F. Disk-ring deposition in drying a sessile
  nanofluid droplet with enhanced marangoni effect and particle surface
  adsorption. \emph{Langmuir} \textbf{2020}, \emph{36}, 15064--15074\relax
\mciteBstWouldAddEndPuncttrue
\mciteSetBstMidEndSepPunct{\mcitedefaultmidpunct}
{\mcitedefaultendpunct}{\mcitedefaultseppunct}\relax
\EndOfBibitem
\bibitem[Chandramohan \latin{et~al.}(2016)Chandramohan, Dash, Weibel, Chen, and
  Garimella]{chandramohan16}
Chandramohan,~A.; Dash,~S.; Weibel,~J.~A.; Chen,~X.; Garimella,~S.~V. Marangoni
  convection in evaporating organic liquid droplets on a nonwetting substrate.
  \emph{Langmuir} \textbf{2016}, \emph{32}, 4729--4735\relax
\mciteBstWouldAddEndPuncttrue
\mciteSetBstMidEndSepPunct{\mcitedefaultmidpunct}
{\mcitedefaultendpunct}{\mcitedefaultseppunct}\relax
\EndOfBibitem
\bibitem[Gelderblom \latin{et~al.}(2022)Gelderblom, Diddens, and
  Marin]{gelderblom22}
Gelderblom,~H.; Diddens,~C.; Marin,~A. Evaporation-driven liquid flow in
  sessile droplets. \emph{Soft Matter} \textbf{2022}, \emph{18},
  8535--8553\relax
\mciteBstWouldAddEndPuncttrue
\mciteSetBstMidEndSepPunct{\mcitedefaultmidpunct}
{\mcitedefaultendpunct}{\mcitedefaultseppunct}\relax
\EndOfBibitem
\bibitem[Kulinich \latin{et~al.}(2011)Kulinich, Farhadi, Nose, and
  Du]{kulinich11}
Kulinich,~S.~A.; Farhadi,~S.; Nose,~K.; Du,~X.~W. Superhydrophobic surfaces:
  are they really ice-repellent? \emph{Langmuir} \textbf{2011}, \emph{27},
  25--29\relax
\mciteBstWouldAddEndPuncttrue
\mciteSetBstMidEndSepPunct{\mcitedefaultmidpunct}
{\mcitedefaultendpunct}{\mcitedefaultseppunct}\relax
\EndOfBibitem
\bibitem[Du \latin{et~al.}(2022)Du, Zhang, and Shen]{du22}
Du,~F.; Zhang,~L.; Shen,~W. The internal flow in an evaporating human blood
  plasma drop. \emph{J. Colloid Interface Sci.} \textbf{2022}, \emph{609},
  170--178\relax
\mciteBstWouldAddEndPuncttrue
\mciteSetBstMidEndSepPunct{\mcitedefaultmidpunct}
{\mcitedefaultendpunct}{\mcitedefaultseppunct}\relax
\EndOfBibitem
\bibitem[Schatz and Neitzel(2001)Schatz, and Neitzel]{schatz01}
Schatz,~M.~F.; Neitzel,~G.~P. Experiments on thermocapillary instabilities.
  \emph{Annu. Rev. Fluid Mech} \textbf{2001}, \emph{33}, :93–127\relax
\mciteBstWouldAddEndPuncttrue
\mciteSetBstMidEndSepPunct{\mcitedefaultmidpunct}
{\mcitedefaultendpunct}{\mcitedefaultseppunct}\relax
\EndOfBibitem
\bibitem[Mizev and Schwabe(2009)Mizev, and Schwabe]{mizev09}
Mizev,~A.~I.; Schwabe,~D. Convective instabilities in liquid layers with free
  upper surface under the action of an inclined temperature gradient.
  \emph{Phys. Fluids} \textbf{2009}, \emph{21}, 112102\relax
\mciteBstWouldAddEndPuncttrue
\mciteSetBstMidEndSepPunct{\mcitedefaultmidpunct}
{\mcitedefaultendpunct}{\mcitedefaultseppunct}\relax
\EndOfBibitem
\end{mcitethebibliography}


\begin{tocentry}
\centering
\includegraphics{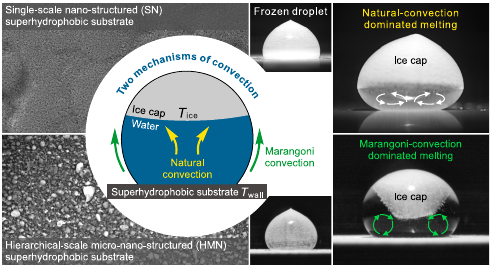}\\
\vspace{3mm}
\end{tocentry}

\end{document}